\documentclass[pre,aps,superscriptaddress,twocolumn,floatfix]{revtex4-1}

\usepackage{dsfont}
\usepackage{mathtext}

\usepackage{mathtools}
\usepackage[usenames,dvipsnames]{xcolor}
\usepackage{amsmath}
\usepackage{amssymb,mathrsfs}
\usepackage{graphicx}
\usepackage{epstopdf}

\usepackage{amsmath,mathrsfs}
\usepackage[dvipsnames]{xcolor}

\newcommand{\prt}{\partial}
\newcommand{\la}{\lambda}

\newcommand{\sn}{\mathrm{sn}}
\newcommand{\cn}{\mathrm{cn}}

\begin{document}
	
	\title{Trigonometric shock waves for the Kaup-Boussinesq system}

	\author{Sergey~K.~Ivanov}
	\affiliation{Moscow Institute of Physics and Technology, Institutsky lane 9, Dolgoprudny, Moscow region, 141700, Russia}
	\affiliation{Institute of Spectroscopy, Russian Academy of Sciences, Troitsk, Moscow, 108840, Russia}
	
	\author{Anatoly~M.~Kamchatnov}
	\affiliation{Moscow Institute of Physics and Technology, Institutsky lane 9, Dolgoprudny, Moscow region, 141700, Russia}
	\affiliation{Institute of Spectroscopy, Russian Academy of Sciences, Troitsk, Moscow, 108840, Russia}

\begin{abstract}
We consider the modulationally stable version of the Kaup-Boussinesq system
which models propagation of nonlinear waves in various physical systems. It is shown that the Whitham
modulation equations for this model have a new type of solutions which describe
trigonometric shock waves. In the Gurevich-Pitaevskii problem of evolution of an
initial discontinuity these solutions correspond to a non-zero wave excitation on
one of the sides of the discontinuity. Our analytical results are confirmed by
numerical calculations.
\end{abstract}

	\maketitle
	
\section{Introduction}\label{sec.introduction}

Dispersion in nonlinear systems can dramatically affect wave profile leading to a host
of new physical wave structures such as solitons and dispersive shock waves. In particular,
it is now well known that in such systems a typical evolution of an initial pulse with
a fairly smooth and large initial profile is accompanied by a gradual steepening followed
by the wave breaking and formation of dispersive shock wave (DSW). Theoretically,
dispersive shock waves, also called undular bores in fluid mechanics applications,
are represented as modulated nonlinear periodic waves and then the process of their
formation and evolution is described in the Gurevich-Pitaevskii approximation
\cite{GurevichPitaevskii-73} by the Whitham theory of modulations \cite{Whitham-65}
(for reviews see~\cite{ElHoefer-16,Kamchatnov-21}). The original formulation of
Gurevich and Pitaevskii approach was applied to description of expanding collisionless shocks
(plasma analogs of DSWs) in framework of the Whitham-averaged equations for the integrable
Korteweg-de Vries (KdV) equation~\cite{KortewegVries-95,BenjaminLighthill-54}.
Due to universality of the KdV equation, this approach can naturally be applied
to many other physical situations and it was extended to many other nonlinear wave
equations. For example, when the condition of unidirectional
propagation of the KdV approximation is relaxed, shallow water waves are described by
various forms of the Boussinesq equations~\cite{Boussinesq-77}. The most convenient
for our purposes form has been derived by Kaup~\cite{Kaup-75}; this is the so-called
Kaup-Boussinesq (KB) system which is also integrable by the inverse scattering transform
method. Periodic solutions of the KB system were obtained in Ref.~\cite{MatveevYavor-79}
and the corresponding Gurevich-Pitaevskii theory was extended to the KB case in
Refs.~\cite{ElGrimshawPavlov-01,ElGrimshawKamchatnov-05,CongyIvanovKamchatnovPavloff-17}.

In the applications of Gurevich-Pitaevskii approach to concrete water wave problems,
the KB equations with negative dispersion are used.
However, the dispersion relation for this kind of equations corresponds to a dynamical
instability of small wavelength perturbations over a fluid of constant depth $h_0$.
There exists another form of the KB system with positive dispersion
\begin{equation}\label{KBeq}
\begin{array}{l}
h_t+(hu)_x-\frac14u_{xxx}=0,\\
u_t+uu_x+h_x=0,
\end{array}
\end{equation}
where $h$ is  the local height of the water layer and $u$ is a local mean
flow velocity. For this equation the dispersion relation of linear waves reads
\begin{equation}\label{eq2}
\omega^2=h_0k^2+\frac14k^4,
\end{equation}
and the system~(\ref{KBeq}) does not suffer from this kind of deficiency. Moreover,
it appears as an approximation to the nonlinear polarization dynamics of a two-component
Bose-Einstein condensate \cite{IvanovKamchatnovCongyPavloff-17} as well as to dynamics of
magnetization in magnetics with easy-plane anisotropy, so it deserves 
thorough investigation. 

In Ref.~\cite{CongyIvanovKamchatnovPavloff-17} the Riemann problem
of evolution of initial discontinuities was studied for the system (\ref{KBeq}).
Here we consider the initial states of a different type: we assume that on one side of the
initial discontinuity the profiles are represented by periodic solutions rather than by
uniform distributions, as it supposed in the standard Riemann problem. This means that
our theory describes spreading out of the front of the nonlinear wave excitation along
the rarefaction wave. This opens a new route to analytical description of wave
structures arising from more complex initial states.

\section{Periodic solutions and Whitham equations}\label{sec.whitham}

In this section, we derive the periodic wave solutions of the system~(\ref{KBeq}) and
the Whitham equations governing the modulational dynamics. The KB system~(\ref{KBeq})
is completely integrable and it can be represented as a compatibility condition of
two linear equations (see~\cite{Kaup-75}) with a free spectral parameter $\la$:
\begin{equation}\label{eq5}
\psi _{xx}=\mathcal{A}\, \psi ,
\qquad
\psi _{t}=-\frac12\mathcal{B}_x\psi+\mathcal{B}\,\psi_x,
\end{equation}
where
\begin{equation}\label{eq6}
\mathcal{A}=h-\left(\la-\frac12 u\right)^2,\quad\mbox{and}\quad
\mathcal{B}=-\left(\la+\frac12 u\right).
\end{equation}
This permits one to find its periodic solutions and Whitham modulation equations.

Periodic solution can be obtained by the well-known finite-gap integration method
(see, e.g.,~\cite{Kamchatnov-2000}).  The second order spatial linear differential
equation in (\ref{eq5})
has two basis solutions  $\psi _{+}(x,t)$ and $\psi_{-}(x,t)$.
Their product $g=\psi _{+}\psi_{-} $ satisfies the
following third order equation
\begin{equation}\label{eq7}
g_{xxx}-2\mathcal{A}_xg-4\mathcal{A}\, g_x=0.
\end{equation}
Upon multiplication by $g$, this equation can be integrated once to
give
\begin{equation}\label{eq8}
\frac12 gg_{xx}-\frac14 g_x^2-\mathcal{A}g^2=P(\la),
\end{equation}
where the integration constant has been written as $P(\lambda )$ since
it can only depend on $\lambda$.  The time dependence of $g(x,t)$ is
determined by the equation
\begin{equation}\label{eq9}
g_{t}=\mathcal{B}\,g_x-\mathcal{B}_xg.
\end{equation}
We are interested in the one-phase periodic solutions of the system
(\ref{KBeq}). They are distinguished by the condition that $P(\lambda
)$ in (\ref{eq8}) is a fourth degree polynomial of the
form
\begin{equation}\label{eq10}
P(\lambda )=\prod_{i=1}^{4}(\lambda -\lambda _{i})=\lambda^{4}-
s_{1}\lambda^{3}+s_{2}\lambda ^{2}-s_{3}\lambda +s_{4}.
\end{equation}
Here $s_i$ are standard symmetric functions of four zeros $\lambda_i$
of the polynomial,
\begin{equation}\label{}
\begin{split}
& s_1=\sum_i\lambda_i,\quad s_2=\sum_{i<j}\lambda_i\lambda_j,
\quad s_3=\sum_{i<j<k}\lambda_i\lambda_j\lambda_k, \\
& s_4=\lambda_1\lambda_2\lambda_3\lambda_4.
\end{split}
\end{equation}
We shall assume that $\la_i$ are ordered according to inequalities
\begin{equation}\label{eq10a}
\la_1\leq\la_2\leq\la_3\leq\la_4.
\end{equation}
Then we find from Eq.~(\ref{eq8}) that $g(x,t)$ is a first-degree
polynomial in $\lambda$, 
\begin{equation}\label{eq11}
g(x,t)=\lambda -\mu (x,t),
\end{equation}
where $\mu (x,t)$ is connected with $u(x,t)$ and $h(x,t)$ by the
relations
\begin{equation}\label{eq12}
\begin{split}
& u(x,t)=s_{1}-2\, \mu (x,t),\\
& h(x,t)=\frac{1}{4}s_{1}^{2}-s_{2}-2\mu^{2}(x,t)
+s_{1}\mu(x,t) ,\end{split}
\end{equation}
which follow from a comparison of the coefficients of like
powers of $\lambda$ on both sides of Eq.~(\ref{eq8}). The spectral
parameter $\lambda $ is arbitrary and on substitution of $\lambda =\mu$
into Eq.~(\ref{eq8}) we obtain the equation for $\mu $,
$$
\mu_{x}=2\sqrt{- P(\mu )},
$$
while a similar substitution into Eq.~(\ref{eq9}) gives
$$
\mu_{t}=-\left(\mu +\frac{1}{2}u\right)\mu _{x}=-\frac{1}{2}s_{1}\mu _{x}.
$$
This shows that $\mu$ and hence $h$ and $u$ depend on the variable
\begin{equation}\label{eq13}
\theta =x-\frac{1}{2}s_{1}t,
\end{equation}
only, so that
\begin{equation}\label{eq14}
V=\frac12 s_1=\frac12\sum_{i=1}^4\la_i
\end{equation}
is the phase velocity of the nonlinear wave,
and $\mu (\theta )$ is determined by the equation
\begin{equation}\label{eq15}
\mu _{\theta }=2\sqrt{ -P(\mu )}.
\end{equation}

\subsection{Periodic solutions}

The real solution of Eq.~(\ref{eq15}) corresponds to oscillations of $\mu$ in one of two
possible intervals,
$\la_1\leq\mu\leq\la_2$
or $\la_3\leq\mu\leq\la_4$.

$\bullet$  We first consider the periodic solution corresponding to oscillations
of $\mu$ in the interval
\begin{equation}\label{eq17}
\la_1\leq\mu\leq\la_2.
\end{equation}
Standard calculation yields, after some algebra, the solution in terms of
Jacobi elliptic functions with the initial condition
$\mu(0)=\la_1$:
\begin{equation}\label{eq18}
\mu(\theta)=\la_2-
\frac{(\la_2-\la_1)
	\cn^2\left(W,m\right)}
{1+\frac{\la_2-\la_1}{\la_4-\la_2}
	\sn^2\left(W, m \right)},
\end{equation}
where $W=\sqrt{(\la_3-\la_1)(\la_4-\la_2)}\,\theta$ and
\begin{equation}\label{eq19}
m=\frac{(\la_2-\la_1)(\la_4-\la_3)}{(\la_3-\la_1)(\la_4-\la_2)}
\end{equation}
Substitution of (\ref{eq18}) into (\ref{eq12}) gives the corresponding
expressions for $u(\theta)$ and $h(\theta)$ for a one-phase periodic
nonlinear wave. The wavelength of the wave is equal to
\begin{equation}\label{eq20}
L=\int_{\la_1}^{\la_2}\frac{d\mu}{\sqrt{-P(\mu)}}=
\frac{2K(m)}{\sqrt{(\la_3-\la_1)(\la_4-\la_2)}},
\end{equation}
where $K(m)$ is the complete elliptic integral of the first
kind~\cite{AbramowitzStegun-1972}.

The soliton solution corresponds to the limit $\la_3\to\la_2$
$(m\to1)$ and in the limit
$\la_2\to\la_1$ we get a small-amplitude harmonic wave
(see~\cite{CongyIvanovKamchatnovPavloff-17}).
However, here we are interested in the solution in the form of a trigonometric wave:
if $\la_4=\la_3$ but $\la_1\neq\la_2$, then we have $m=0$ and
Eq.~(\ref{eq18}) reduces to a nonlinear trigonometric wave
\begin{equation}\label{eq21}
\begin{split}
\mu(\theta)&=\la_2-\frac{(\la_2-\la_1)\cos^2W}{1+\frac{\la_2-\la_1}{\la_4-\la_2}\sin^2W},\\
W&=\sqrt{(\la_4-\la_1)(\la_4-\la_2)}\,\theta.
\end{split}
\end{equation}
If we take the limit $\la_2-\la_1\ll \la_4-\la_1$ in this solution, then we
get the small-amplitude limit
\begin{equation}\label{equ25}
\begin{split}
\mu(\theta)\cong \la_2-\frac12(\la_2-\la_1)\cos[(\la_3-\la_1)\theta].
\end{split}
\end{equation}
On the other hand, if we take here the limit $\la_2\to \la_3=\la_4$, then the
argument of the trigonometric functions becomes small and we can
approximate them by the first terms of their series expansions. This
corresponds to an algebraic soliton of the form
\begin{equation}\label{eq21a}
\mu(\theta)=\la_2-\frac{\la_2-\la_1}{1+(\la_2-\la_1)^2\theta^2}.
\end{equation}

$\bullet$ In a similar way, in the second case,
\begin{equation}\label{eq22a}
\la_3\leq\mu\leq\la_4,
\end{equation}
the cnoidal wave solutions are given by the expressions ($\mu(0)=\la_4$)
\begin{equation}\label{eq23}
\mu(\theta)=\la_3+\frac{(\la_4-\la_3)
	\cn^2\left(W,m\right)}
{1+\frac{\la_4-\la_3}{\la_3-\la_1}
	\sn^2\left(W, m \right)}.
\end{equation}
Here the soliton limit can be obtained at $\la_3\to\la_2$ $(m\to1)$.
In the limit
$\la_4\to\la_3$ we get a small-amplitude harmonic wave.
In this case, nonlinear trigonometric waves also exists
and is attained for $\la_1=\la_2$ (but $\la_3\neq\la_4$)
\begin{equation}\label{eq24}
\begin{split}
\mu(\theta)&=\la_3+\frac{(\la_4-\la_3)\cos^2W}{1+\frac{\la_4-\la_3}{\la_3-\la_1}\sin^2W},\\
W&=\sqrt{(\la_3-\la_1)(\la_4-\la_1)}\,\theta.
\end{split}
\end{equation}
If furthermore $\la_3\to\la_1$, then one gets the algebraic
soliton:
\begin{equation}\label{alg-sol}
\mu(\theta)=\la_1+\frac{\la_4-\la_1}{1+(\la_4-\la_1)^2\theta^2}\; .
\end{equation}
At last, if we assume that $\la_4-\la_3\ll\la_4-\la_1$, then we arrive at
the small-amplitude limit
\begin{equation}\label{equ24}
\begin{split}
\mu(\theta)\cong \la_3-\frac12(\la_4-\la_3)\cos[(\la_3-\la_1)\theta].
\end{split}
\end{equation}

\subsection{Whitham modulation equations}

Now we shall consider slowly modulated waves. In
this case, the parameters $\la_i$ ($i = 1$, $2$, $3$, $4$) become slowly
varying functions of $x$ and $t$ changing little in one period
and they can serve as Riemann invariants of modulation equations 
(see~\cite{Kamchatnov-2000}). Evolution of $\la_i$
is governed by the Whitham equations
\begin{equation}\label{eq26}
\frac{\prt\la_i}{\prt t}+v_i\, \frac{\prt\la_i}{\prt x}=0,
\quad i=1,2,3,4.
\end{equation}
The  Whitham velocities $v_i$ can be computed by means of the formulas
\begin{equation}\label{eq27}
v_i(\la_1,\la_2,\la_3,\la_4)
=\left(1-\frac{L}{\partial_{\lambda_i}L}\prt_{\lambda_i}\right)V,
\quad i=1,2,3,4,
\end{equation}
where the phase velocity $V$ and the wavelength $L$ are given
by Eqs.~(\ref{eq14}) and~(\ref{eq20}). A simple
calculation yields the explicit expressions
\begin{equation}\label{eq28}
\begin{array}{l}
\displaystyle{
	v_1=\frac12\sum_{i=1}^4\la_i-\frac{(\la_4-\la_1)(\la_2-\la_1)K(m)}
	{(\la_4-\la_1)K(m)-(\la_4-\la_2)E(m)},}\\[5mm]
\displaystyle{
	v_2=\frac12\sum_{i=1}^4\la_i+\frac{(\la_3-\la_2)(\la_2-\la_1)K(m)}
	{(\la_3-\la_2)K(m)-(\la_3-\la_1)E(m)},}\\[5mm]
\displaystyle{
	v_3=\frac12\sum_{i=1}^4\la_i-\frac{(\la_4-\la_3)(\la_3-\la_2)K(m)}
	{(\la_3-\la_2)K(m)-(\la_4-\la_2)E(m)},}\\[5mm]
\displaystyle{
	v_4=\frac12\sum_{i=1}^4\la_i+\frac{(\la_4-\la_2)(\la_4-\la_1)K(m)}
	{(\la_4-\la_1)K(m)-(\la_3-\la_1)E(m)},}
\end{array}
\end{equation}
where $m$ is given by (\ref{eq19}) and
$K(m)$ and $E(m)$ are complete elliptic integrals of
the first and second kind, respectively.

In the limit $\la_2\to\la_1$ (i.e.,
$m\to0$) we obtain
\begin{equation}\label{eq30}
\begin{split}
& v_1=v_2=2\la_1+\frac{(\la_4-\la_3)^2}{2(\la_3+\la_4-2\la_1)},\\
& v_3=\frac12(3\la_3+\la_4),\quad v_4=\frac12(\la_3+3\la_4),
\end{split}
\end{equation}
and in another limit $m\to0$, i.e. $\la_3\to\la_4$, we have
\begin{equation}\label{eq31}
\begin{split}
&    v_1=\frac12(3\la_1+\la_2),\quad v_2=\frac12(\la_1+3\la_2),\\
&    v_3=v_4=2\la_4+\frac{(\la_2-\la_1)^2}{2(\la_1+\la_2-2\la_4)}.
\end{split}
\end{equation}

Having received the basic equations, we can now proceed to the description
of nonlinear trigonometric solutions for the KB system (\ref{KBeq}).

\section{Trigonometric shock wave}\label{sec.TSW}

\begin{figure}[t]
	\begin{center}
\includegraphics[width=0.7\linewidth]{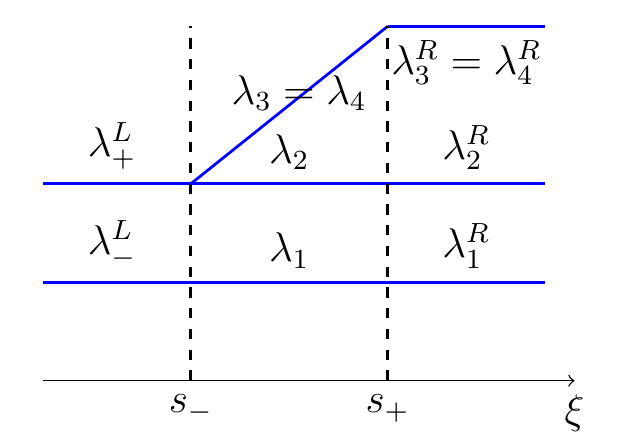}
		\caption{Sketches of the behavior of the Riemann invariants
			in trigonometric dispersive shock wave solutions of the Whitham
			equations with $\la_3=\la_4$. Vertical dashed lines indicated by $s_-$ and $s_+$ define the edges of a trigonometric shock wave. Corresponding wave structure is shown
in Fig.~\ref{fig2}.}
		\label{fig1}
	\end{center}
\end{figure}

\begin{figure*}[t]
	\begin{center}
		\includegraphics[width=0.9\linewidth]{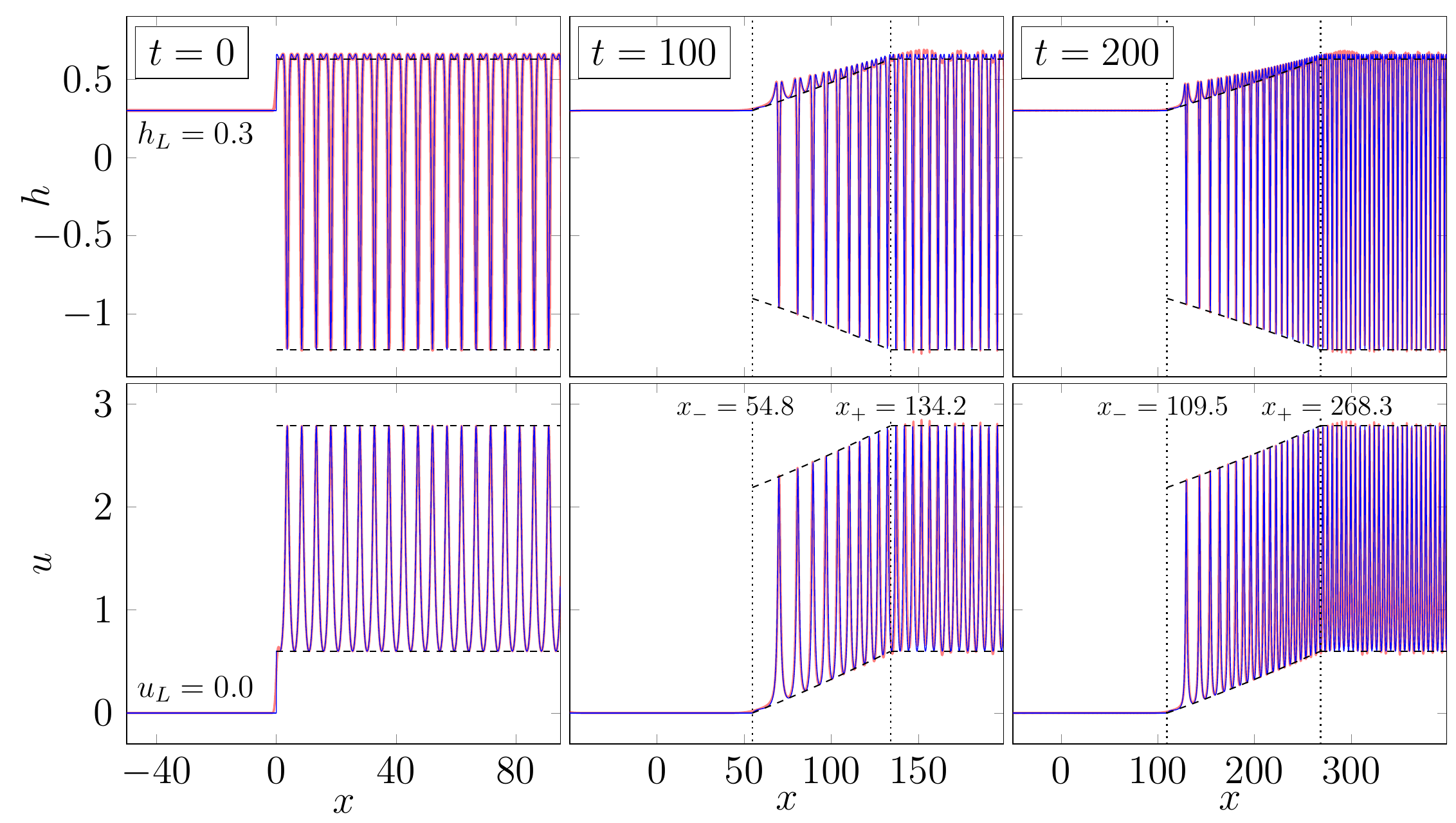}
		\caption{Evolution of a trigonometric shock wave for $h_L=0.3$,
$u_L=0$ and $\lambda_3^R=\lambda_4^R=0.85$. Figures show the initial state
(left column) and wave profiles for depth $h$ and flow velocity $u$ for $t=100$
(middle column) and $t=200$ (right column). Red (thick) curves show the result
of numerical calculations, and a blue (thin) one shows the analytical solution.
Dashed lines illustrate envelopes of wave structure, vertical dashed lines indicate
the edges of the trigonometric shock wave ($x_-$ and $x_+$). We have here dark
solitons of elevation $h$ and bright solitons of flow velocity $u$ at the soliton edge of
the shock. The corresponding diagram of Riemann invariants is shown in Fig.~\ref{fig1}. }
		\label{fig2}
	\end{center}
\end{figure*}

\begin{figure}[t]
	\begin{center}
		\includegraphics[width=0.7\linewidth]{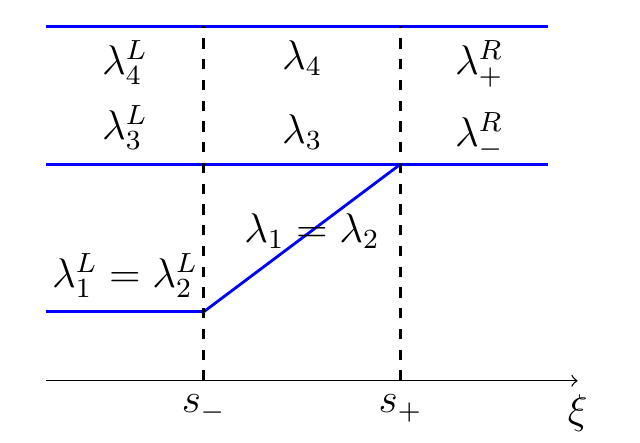}
		\caption{Sketches of the behavior of the Riemann invariants
			in trigonometric dispersive shock wave solutions of the Whitham
			equations with $\la_1=\la_2$. Vertical dashed lines indicated by $s_-$ and $s_+$ define the edges of a trigonometric shock wave. Corresponding wave structure
is shown in Fig.~\ref{fig4}.}
		\label{fig3}
	\end{center}
\end{figure}

\begin{figure*}[t]
	\begin{center}
		\includegraphics[width=0.9\linewidth]{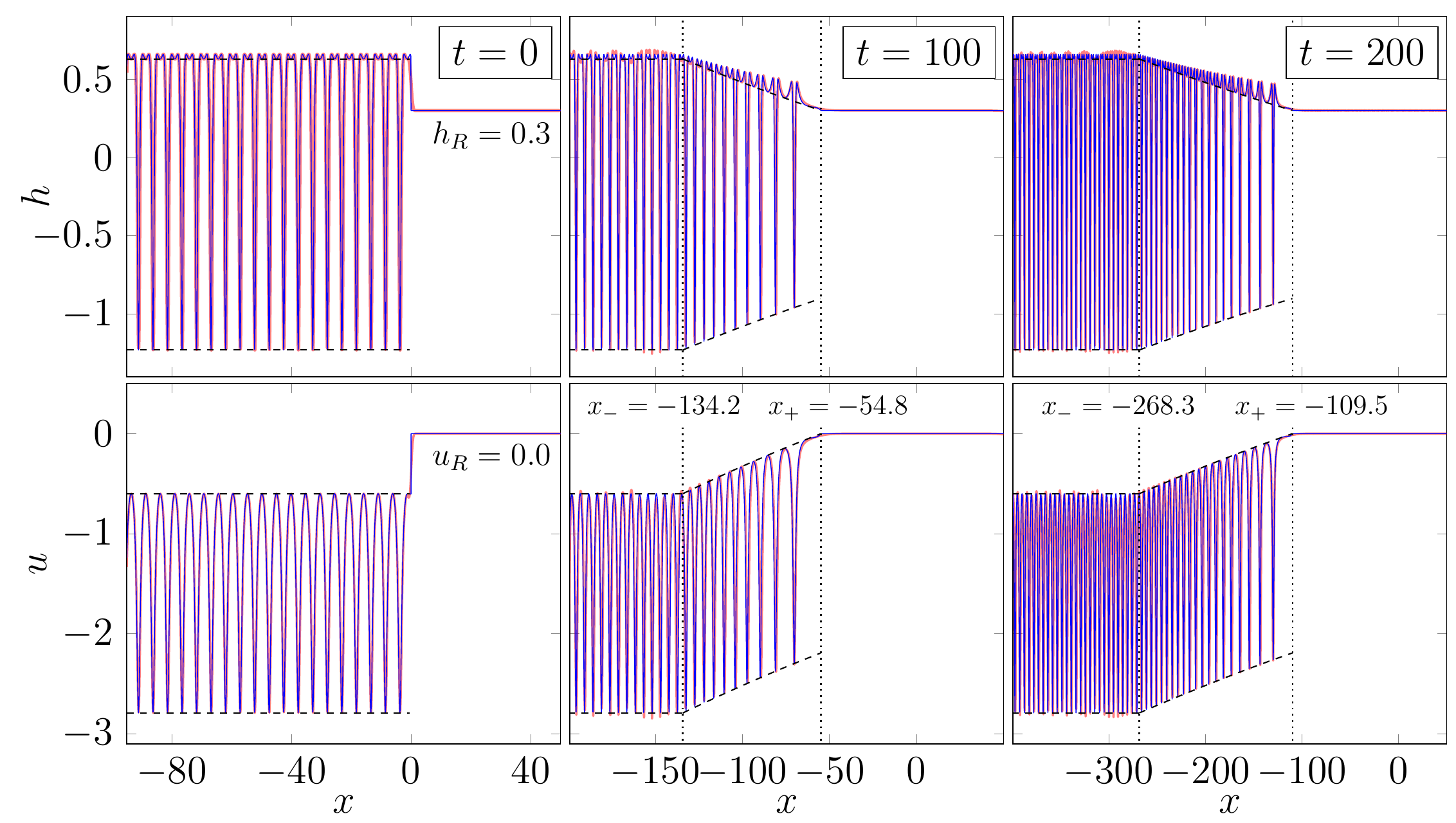}
		\caption{Evolution of a trigonometric shock wave $h_R=0.3$, $u_R=0$
and $\lambda_1^L=\lambda_2^L=-0.85$. Figures show the initial state (left column)
and wave profiles for depth $h$ and flow velocity $u$ for $t=100$ (middle column)
and $t=200$ (right column). Red (thick) curves show the result of numerical
calculations, and a blue (thin) one shows the analytical solution. Dashed lines
illustrate envelopes of wave structure, vertical dashed lines indicate the edges
of the trigonometric shock wave ($x_-$ and $x_+$). On the contrary to the
case of Fig.~\ref{fig2}, now we get dark
solitons of both elevation $h$ and  flow velocity $u$ at the soliton edge of
the shock. The corresponding diagram of
Riemann invariants is shown in Fig.~\ref{fig3}.}
		\label{fig4}
	\end{center}
\end{figure*}

Here we consider formation of the trigonometric shock wave pattern. In case of
conventional initial conditions with uniform distributions on both sides of the
initial step-like discontinuity such a structure is not generated
\cite{CongyIvanovKamchatnovPavloff-17}. However, there exist quite natural initial
conditions for which this kind of solutions does appear.
Let the initial conditions have the following form. On the one side of the point
$x = 0$ we have constant distributions of $h$ and $u$, and on the other side there is
a non-modulated nonlinear periodic wave described by four constant parameters
$\lambda_i$. This means that we are interested in evolution of an initially sharp
border of the wavy region. Before turning to oscillatory solutions, let us first
consider the dispersionless case.

For smooth enough wave distributions we can neglect the last dispersion term in
the first equation of the system (\ref{KBeq}) and arrive at the
dispersionless equations
\begin{equation}\label{eq29a}
h_t+(hu)_x=0,\qquad u_t+uu_x+h_x=0,
\end{equation}
which coincide with the well-known shallow water equations. First of all, this system
admits a trivial solution for which $h = \mathrm{const}$ and $u = \mathrm{const}$.
We shall call such a solution a ``plateau''. We introduce the so-called Riemann invariants
\begin{equation}\label{eq31a}
\la_{\pm}=\frac{u}2\pm\sqrt{h}\;.
\end{equation}
Using these dispersionless Riemann invariants, equations \eqref{eq29a} can be
written in the following diagonal form
\begin{equation}\label{eq30a}
\begin{split}
\frac{\prt\la_{\pm}}{\prt t}+v_\pm(\la_-,\la_+)\frac{\prt\la_\pm}{\prt x}=0,
\end{split}
\end{equation}
where
\begin{equation}\label{eq30b}
\begin{split}
v_\pm(\la_-,\la_+)=\frac12(3\la_\pm+\la_\mp)\; .
\end{split}
\end{equation}
These dispersionless variables are required for the correct choice of the initial state.

We assume that at the initial moment the profile is divided into two parts by the point $x = 0$.
We consider two types of profile. First, at $x <0$ there is a plateau characterized
by the constant dispersionless invariants $\la_-=\mathrm{const}$, $\la_+=\mathrm{const}$,
and for $x> 0$ there is a non-modulated wave described by the Eq~(\ref{eq21})
(see Fig.~\ref{fig1} and left column in Fig.~\ref{fig2}). We shall denote the parameters
of this periodic wave as $\la_1^R$,  $\la_2^R$, and  $\la_3^R=\la_4^R$. The second type of
initial state is similar: there is a non-modulated wave described by Eq~(\ref{eq24})
with $\la_1^L=\la_2^L=\mathrm{const}$, $\la_3^L=\mathrm{const}$,  $\la_4^L=\mathrm{const}$
at $x <0$ and a plateau with $\la_-=\mathrm{const}$, $\la_+=\mathrm{const}$ at $x>0$
(see Fig.~\ref{fig3} and left column in Fig.~\ref{fig4}).

We now turn to the study of situations when the dispersion effects are taken into account. In order to
satisfy the matching conditions, the only possible solution arising from a given initial
state may be a trigonometric shock wave. Trigonometric shock wave can be represented
approximately as a modulated nonlinear periodic wave in which parameters $\lambda_i$
change slowly along the wave structure. In such a modulated periodic solution two equal Riemann invariants are changing and the other two remain constant along the entire shock.
Thus, in our case the constant Riemann invariants and the Riemann invariants at the boundaries
of the structure should have equal values. This situation resembles the so called
`contact discontinuity' which plays an important role in the theory of viscous shocks
(see, e.g., ~\cite{LL-6}). This type of DSW was first reported in Ref.~\cite{Marchant-08} where
the evolution of a step problem was studied for the focusing modified KdV equation.
In Ref.~\cite{ElHoeferShearer-17} these (trigonometric) DSWs were called contact DSWs.
The trigonometric shock waves for the KB system are described by the modulated finite-amplitude
nonlinear periodic solutions (\ref{eq21}) or (\ref{eq24}). The evolution of the trigonometric
shock wave is determined by the Whitham equations (\ref{eq26}). In our case of the step-like
initial conditions we have to find self-similar solutions for which all Riemann invariants
depend only on $\xi = x/t$, and the Whitham equations reduce to
\begin{equation}\label{eq35}
\frac{d\la_i}{d\xi}\cdot
\left[v_i(\la_1,\la_2,\la_3,\la_4)-\xi\right]=0,\quad i=1,2,3,4.
\end{equation}
If, for instance, we consider the first type of initial condition then in this situation,
shown in Fig.~\ref{fig1}, trigonometric shock wave has two equal parameters
$\lambda_3=\lambda_4$ and the invariants $\la_1$ and $\la_2$ are constant along
the whole wave pattern including shock region. We assume that there is a plateau with
dispersionless Riemann $\la_-^L$ and $\la_+^L$ invariants to the left of the
trigonometric shock wave. Thus we have
$v_3(\la_-^L,\la_+^L, \lambda_4(\xi),\lambda_4(\xi)) = v_4(\la_-^L,\la_+^L,
\lambda_4(\xi),\lambda_4(\xi))=\xi$. Consequently, we obtain
\begin{equation}\label{}
\begin{split}
& \lambda_1=\lambda_-^L, \quad \lambda_2=\lambda_+^L,\\
& v_4=2\lambda_4-\frac{(\lambda_+^L-\lambda_-^L)^2}{\lambda_+^L+\lambda_-^L-2\lambda_4}=\xi,
\end{split}
\end{equation}
where the last formula determines the dependence of $\lambda_4$ on $\xi$,
which can be represented in the explicit form
\begin{equation}\label{}
\begin{split}
\la_4(\xi)=\frac14
& \bigg[\xi+\la_+^L+\la_-^L+\\
& \sqrt{(\xi-\la_+^L-\la_-^L)^2+2(\la_+^L-\la_-^L)^2}\bigg].
\end{split}
\end{equation}
Here $\xi$ varies within the interval $s_-\leq \xi\leq s_+$ with
\begin{equation}\label{}
\begin{split}
s_-=\frac{3\la_+^L+\la_-^L}{2},\quad
s_+=2\lambda_4^R+\frac{(\la_+^L-\la_-^L)^2}{2(\la_+^L+\la_-^L-2)},
\end{split}
\end{equation}
where $\la_4^R=\la_3^R$ is the maximum value of $\la_4(\xi)$ defined
by the right boundary condition of trigonometric shock wave.
The wavelength in this case is given by the formula
\begin{equation}\label{}
L=\frac{2\pi}{\sqrt{(\la_4(\xi)-\la_-^L)(\la_4(\xi)-\la_+^L)}}.
\end{equation}

Substitution of $\la_i$ into Eq.~(\ref{eq21}) with subsequent substitution
into Eq.~(\ref{eq12}) yields the modulated periodic solutions resulting
in the trigonometric shock wave structure. Comparison of the obtained analytical
solution with numerical calculations is shown in Fig.~\ref{fig2} for different
values of time $t$.
One can see that the trigonometric shock wave is located between the edges with
coordinates $x_-=s_-t$ and $x_+=s_+t$. This wave matches at its left edge with
the left  plateau and at its right edge with the non-modulated wave.

In a similar way, we can consider the second type of initial conditions.
For this case the trigonometric shock wave has $\la_1=\la_2$. An example of
the diagram of Riemann invariants is shown in Fig.~\ref{fig3}. Obviously,
we get a symmetric situation, where the trigonometric shock wave matches the plateau
characterised by the values $\la_-^R$ and $\la_+^R$ of the Riemann invariants 
at the right edge and the non-modulated
wave with the values $\la_1^L=\la_2^L=\mathrm{const}$, $\la_3^L=\la_-^R$ and $\la_2^L=\la_+^R$
of the Riemann invariants of the Whitham system
at the left edge. The solution of the Whitham equations takes the form
\begin{equation}\label{284.2}
\begin{split}
& v_1=v_2=2\la_1+\frac{(\la_+^R-\la_-^R)^2}{2(\la_+^R+\la_-^R-2\la_1)}=\xi,\\
& \la_3=\la_-^R=\la_-^R,\quad \la_4=\la_+^R=\la_+^R,
\end{split}
\end{equation}
or
\begin{equation}\label{285.2}
\begin{split}
\la_1(\xi)=\frac14 &
\bigg[\xi+\la_+^R+\la_-^R-\\
& \sqrt{(\xi-\la_+^R-\la_-^R)^2+2(\la_+^R-\la_-^R)^2}\bigg]
\end{split}
\end{equation}
where $\xi$ varies in the interval $s_-\leq \xi\leq s_+$ with
\begin{equation}\label{285.3}
s_-=-2\la_1^L+\frac{(\la_+^R-\la_-^R)^2}{2(\la_+^R+\la_-^R+2)},\quad
s_+=\frac{\la_+^R+3\la_-^R}{2}.
\end{equation}
Parameter $\la_1^L=\la_2^L$ is determined again by the initial conditions
at the boundary with the non-modulated wave. The wavelength is given here
by the formula
\begin{equation}\label{si4}
L=\frac{2\pi}{\sqrt{(\la_1(\xi)-\la_-^R)(\la_1(\xi)-\la_+^R)}}.
\end{equation}
Our analytical results and numerical simulations for the corresponding
wave structures are compared in Fig.~\ref{fig4}. Now trigonometric shock wave is
located between the edge points $x_-=s_-t$ and $x_+=s_+t$ indicated by dotted lines.
One can see that numerical calculations (red thick) agree with the analytical
curve (blue thin) very well.

Although the wave patterns look similar, one should notice that in Fig.~\ref{fig2}
the distribution of $u(x,t)$ tends to bright solitons at the soliton edge of the
shock whereas in Fig.~\ref{fig4} we get dark solitons of $u(x,t)$ at this edge.
In both cases the variable $h(x,t)$ takes negative values and that limits application
of the developed theory to water waves. Nevertheless it is applicable to other
physical situations such as nonlinear shocks in two-component Bose-Einstein
condensates or in magnetics.

~

\section{Conclusion}

In this Letter we have considered the cases in which the trigonometric shock waves
arise when their evolution is governed by the integrable Kaup-Boussinesq system.
The initial state contains a nonlinear wave which significantly extends the set
of wave patterns which can be generated in various physical situations. Our results
can find applications as approximation to the dynamics of polarization waves in
two-component Bose-Einstein condensates and in magnetic systems with easy-plane
anisotropy which are not constrained by the condition that $h$ must be positive.

\section{Acknowledgments}

This work was supported by a grant from Foundation for the
Advancement of Theoretical Physics and Mathematics ``BASIS''.

\end{document}